\documentstyle[12pt]{article}

\begin{document}
\def\la{\mathrel{\mathpalette\fun <}}
\def\ga{\mathrel{\mathpalette\fun >}}
\def\fun#1#2{\lower3.6pt\vbox{\baselineskip0pt\lineskip.9pt
        \ialign{$\mathsurround=0pt#1\hfill##\hfil$\crcr#2\crcr\sim\crcr}}}
\newcommand {\eegg}{e^+e^-\gamma\gamma~+\not \! \!{E}_T}
\newcommand {\mumugg}{\mu^+\mu^-\gamma\gamma~+\not \! \!{E}_T}
\renewcommand{\thefootnote}{\fnsymbol{footnote}}
\bibliographystyle{unsrt}

\begin{flushright} \small{
IFUP-TH-32/96\\
UMD-PP-96-108\\
hep-ph/9606408\\
June, 1996}
\end{flushright}

\vspace{2mm}

\begin{center}
 
{\Large \bf A Non-supersymmetric Interpretation of
the CDF $\eegg$ Event}\\

\vspace{1.3cm}
{\large Gautam Bhattacharyya$^{1}$\footnote{gautam@ipifidpt.difi.unipi.it} 
and  
Rabindra N.\ Mohapatra$^{2}$ \footnote{rmohapatra@umdhep.umd.edu}\\    
[5mm]
$^1${\em Dipartimento di Fisica, Universit\`{a} di Pisa\\
and INFN, Sezione di Pisa, I-56126 Pisa, Italy }\\
\vspace{2mm}
$^2${\em Department of Physics, University of Maryland, \\
College Park, Maryland 20742, USA}\\} 
\end{center}

\vspace{2mm}
\begin{abstract}

The $\eegg$ event reported recently by the CDF
Collaboration has been interpreted as a signal of supersymmetry 
in several recent papers. In this article, we report on an alternative
non-supersymmetric interpretation of the event using an extension of
the standard model which contains new physics at the electroweak scale
that does not effect the existing precision electroweak data. We extend
the standard model by including an extra sequential generation of 
fermions, heavy right-handed neutrinos for all generations and an
extra singly charged SU(2)-singlet Higgs boson. We discuss possible
ways to discriminate this from the standard supersymemtric 
interpretations.

\end{abstract}
 
\newpage
\renewcommand{\thefootnote}{\arabic{footnote})}
\setcounter{footnote}{0}
\addtocounter{page}{-1}
\baselineskip=24pt

The Fermilab CDF collaboration\cite{cdf} has recently 
reported an event which contains a hard electron-positron pair 
with two hard photons and missing transverse energy. 
The standard model (SM) background for this event is negligible; 
therefore, if more events like
this are obtained further, it will indeed signal the existence
of new physics beyond the SM. In two recent 
papers \cite{thomas,kane}, it has been proposed that this single event is
consistent with a supersymmetric interpretation when e.g.
$q\bar{q}\rightarrow \tilde{e}_R\tilde{\bar{e}}_R$ with
either (i)
$\tilde{e}_R\rightarrow e+\tilde{\gamma}$ followed by $\tilde{\gamma}
\rightarrow \gamma + \tilde{G}$ 
or (ii) $\tilde{e}_R\rightarrow e + \chi_2$
followed by $\chi_2\rightarrow \chi_1 +\gamma$ 
($\tilde{G}$ denotes a massless goldstino in the gauge mediated low
energy supersymmetry-breaking scenario and $\chi_{1,2}$ denote the 
lightest and the second lightest neutralino respectively).
Clearly, this has given further boost to the activities in the area of
supersymmetry (SUSY) which already enjoys a number of theoretical 
advantages in terms of understanding the puzzles of the SM. While, such
$\eegg$ (or for that matter  $\mumugg$ if they
appear) receives a natural interpretation in terms of SUSY, 
before one can be completely sure about this, one must rule out any other
reasonable non-supersymmetric interpretation. The purpose
of this note is to point out that the reported experimental features of the
single $\eegg$ can be obtained in a simple weak scale extension of the
SM without invoking SUSY. While the model
we present is completely consistent with all known low energy data and could
easily be a viable model of particle physics at the electroweak scale, our 
goal is more to present it as a possible alternative to SUSY that can fake
the CDF signal. If more such `zoo event' accumulate, an experimental 
discrimination is necessary
before one can accept {\em prima facie} that SUSY is manifesting. 

The model we propose is based on the SM gauge group 
$SU(2)_L\times U(1)_Y$. In addition to the particles of the SM, it
contains (i) an extra sequential generation denoted by 
$Q_4\equiv (t',b')_L$, $t'_R,b'_R$, $L_4\equiv (N,E)_L$, $N_R$, $E_R$,
(ii) right-handed $SU(2)$-singlet neutrinos ($\nu_{iR}$) corresponding
to the first three  
generations and (iii) a singly charged $SU(2)$-singlet scalar
denoted by $\eta(\equiv \eta^\pm)$ which can only
couple to $L_4$ and not to $Q_4$. 
It may be noted that 
a heavy sequential generation of degenerate fermions contributes 
$+2/3\pi$ to the oblique electroweak parameter $S$ and with the 
present precision of electroweak data one complete sequential 
generation can still be accommodated \cite{langa}. The fermions of
the fourth generation are kept heavy enough so that they do not effect
any other consequence of the SM. 
The relevant part of the new Yukawa Lagrangian of the model 
looks like
\begin{eqnarray}
L^{new}_{Y} & = & f_i \eta^+l_{iR}\nu_{iR} +
f'_i \eta^+l_{iR}N_R  + f_{4i}\eta^+E_R\nu_{iR} 
+ f_{Ei}\eta^+L_4L_i \nonumber \\
 & + & f_{ij}\eta^+L_i L_j + h L_iH \nu_{iR} + {\rm h.c.}
\end{eqnarray}
where $l_i=e,\mu,\tau$; the subscript $i,j$ also go over $e,\mu,\tau$;
$L_4$ and $L_i$ in the above equation denote the $SU(2)_L$-doublet
part of the fourth- and the first three- generations respectively. 
In the first term in the Lagrangian, we have kept only the diagonal
terms for simplicity. To start with, let us assume that 
$i=e$, i.e. new physics couples only to the first generation, except 
for $f_{ij}$ where antisymmetry in the indices imply $j = \mu$ or
$\tau$.  
$\nu_{iR}$ have large Majorana masses in the $\sim 65$ GeV region.  
The smallness of the left-handed SM neutrino masses can be 
explained by adjusting the off-diagonal Dirac masses invoking the
usual see-saw mechanism. We will show below that if 
$M_{E,N}> M_{\eta} > M_{\nu_{eR}}$ are satisfied 
and if $f_{ij}$ is vanishingly small,
then in a hadron collider, one can pair produce $\eta$ by gauge 
interactions with
$\eta\rightarrow e_R \nu_{eR}$ followed by 
$\nu_{eR}\rightarrow \nu_e +\gamma$.
To explain the kinematics of the $\eegg$ event,
we will assume that $M_{\eta}\simeq 100$ GeV and  $M_{\nu_{eR}}\simeq 
65$ GeV.
We will show that for our choice of the parameters, both the above decays 
constitute almost 100\% branching ratios and the emerging final 
states (electrons and $\gamma$'s) are hard as required. 
It is a necessity to assume the existence of the fourth-generation 
leptons which in conjunction with $f_{ij} = 0$ guarantees a virtually 100\% 
branching ratio to the $\nu_{eR}\rightarrow \nu_e \gamma$ decay mode 
and prevents other channels (such as $\nu_\mu \bar{\nu}_\mu e^+e^-$ etc.) 
from appearing as final states in $\nu_{eR}$ decay. Moreover, the 
coupling $h$ has to be smaller than $\sim 0.1$ to suppress decay
modes like $\nu_{eR} \rightarrow \nu_e b\bar{b}$.   
This non-supersymmetric scenario can provide 
as good an explanation of the CDF $\eegg$ event.

In the simplest version of the model with 
$f_{e}, f_{4e}, f_{Ee} \neq 0$, $f_{ij} = 0$ and all other couplings 
involving the second and third generation leptons switched off, 
the mass hierarchy mentioned in the previous paragraph implies 
that all the new heavy 
particles except the $\nu_{eR}$ have tree-level decays to lighter 
particles by virtue of the interactions in eq. 1. In fact
it is required that all heavy paricles must decay into lighter ones
before $\sim 1$ second or so  
since injecting extra energy at the nucleosynthesis
era is cosmologically troublesome. Guarded by all these requirements 
we are now set to see how this model can explain the $\eegg$ event. 

The first step is the pair production of $\eta$'s by gauge interactions. 
Since the $\eta$ has the
same gauge quantum number as the $\tilde{e}_R$, its production 
cross section is
at the 10 fb level for mass of order 100 GeV or so (see e.g. 
\cite{thomas,kane,babu} for numerical details). 
Being lighter than $E$ or $N$, $\eta$ will decay 
to $\nu_{eR}+ e$ with a strength proportional to $f_e^2$; we assume that
the $M_{\eta}- M_{\nu_{eR}}\simeq 35 $ GeV or so to understand the observed
electron energy. Let us now look for the decay of $\nu_{eR}$; since we set
$f_{ij}=0$ and $M_{\eta}> M_{\nu_{eR}}$, the only tree level decays
for the $\nu_{eR}$ are through its mixings with the light neutrino
via the see-saw mechanism and these decays  can be either $Z$-mediated
or $W^{\pm}$-mediated leading to 
$\nu_{eR}\rightarrow 3\nu$ or $\nu_{eR}\rightarrow
\nu l^+ l'^-$. The decay widths for these processes are given by:
$\Gamma_{ 3\nu~or~\nu e^+e^-}\simeq {{G^2_FM^5_{\nu_{eR}}}\over{192\pi^3}}
\left({{m_{\nu_L}}\over{M_{\nu_{eR}}}}\right)$; note that
they are suppressed by the small neutrino masses. 
However at the one loop level, one gets the penguin decay
 $\nu_{eR}\rightarrow \nu_e +\gamma$. The amplitude for this decay arises
from the $E$ and $\eta$ flowing as virtual particles in 
the loop. This decay is controlled by the heavy fourth generation masses 
and its amplitude is estimated to be 
\begin{equation}
A \left(\nu_{eR} \rightarrow \nu_e\gamma\right) \simeq 
{{f_{4E}f_{Ee} e}\over{16\pi^2 M_E}}
\end{equation}
Although this is a loop decay, it can dominate the tree level decay 
which is suppressed by light neutrino masses, mentioned earlier. The 
one-loop decay width for the $\nu_{eR}$ is about $\Gamma_{\nu_{eR}}
\simeq 1.8\times 10^{-10}$ GeV for 
$f_{4e}\simeq f_{Ee}\simeq 10^{-1}$ for $M_{\nu_{eR}}
\simeq 65 $ GeV and $M_E\simeq 150$ GeV or so. 
Note that the presence of the fourth generation lepton
is crucial for this purpose. The purely $W$- and $Z$-mediated decay
widths mentioned above are much smaller than the
photonic decay mode if $m_{\nu_L}<$ 4.5 KeV for $M_{\nu_{eR}}= 65$
GeV, leading to  $\nu_{eR} \rightarrow \nu_e+\gamma$
as the dominant decay mode of the $\nu_{eR}$.
The kinematics is similar to the gravitino mode discussed
in refs. \cite{thomas,kane}. We also expect the $\nu_{eR}$ to travel about
$\sim 10^{-3}\left(10^{-2}/f_{4e}f_{Ee}\right)^2~mm$ before decay. 
For lower values of the $f$ parameters, one should observe a 
displaced vertex for the photons from the $e^+e^-$.

An interesting set of predictions follow if we switch on the muon 
couplings
in the model (i.e. $f_{\mu}, f_{4\mu}, f_{E\mu}\neq 0$). If we 
assume analogously that $M_{\nu_{\mu R}}< M_{\eta}$, we would expect 
the branching ratio for the electron to muon modes to be proportional
to $f^2_e/f^2_{\mu}$; as a result, one would get also
$\mumugg$-type events in $p\bar{p}$ collider experiments
if the muon-neutrino mass is assumed to be less than 4.5 KeV.

However, the presence of both $f_{Ee}$ and $f_{E\mu}$ will lead to the
rare process like $\mu\rightarrow e\gamma$ or $\mu\rightarrow 3e$. 
This in turn will put constraints
on the simultaneous production of both $ee$- and $\mu\mu$-type events.
To see these constraints in detail, we calculate the 
$B(\mu\rightarrow e+\gamma)$ and find that the present upper limit of
$4.9\times 10^{-11}$ on it implies that 
$f_{Ee}f_{E\mu }< 6 \times 10^{-5}$ and 
$f_{Ee}f'_e< 6 \times 10^{-8}$.
Once $\mu\rightarrow e\gamma$ bound
is satisfied, $\mu\rightarrow 3e$ is also seen to be satisfied.
Requiring the $\nu_{eR}$ and the $\nu_{\mu R}$ to decay inside the
detector puts the following constraints on the couplings: 
$f_{4e}f_{Ee}>8 \times 10^{-6}$ and $f_{4\mu}f_{E\mu} > 8 \times 10^{-6}$. 
It is possible to satisfy
all these constraints simultaneously by appropriately choosing the 
Yukawa coupling parameters. 

In this scenario, one should expect the 
number of events of $ee$-, $\mu\mu$- and $e\mu$-types to satisfy the
relation $ N^2_{e\mu}=N_{ee} N_{\mu\mu}$, which is different from the
prediction of the SUSY model \cite{thomas,kane} where any 
mixed $e\mu$-type
event will arise only from the $\tau\tau$-type events. In our case
number of $\tau\tau$-type events will be proportional to another
parameter $f_{\tau}$ and is therefore arbitrary. The relative number
of $ee\gamma\gamma$- and $\mu\mu\gamma\gamma$-type events can therefore
be used to distinguish this model from its SUSY counterpart.

A few additional comments regarding the model are in order:

\noindent (i)  
The new Yukawa interaction will induce corrections to $Z \rightarrow 
ee, \mu\mu$ and also to $Z \rightarrow inv.$ at the one-loop level via
$\eta$- and $L_4$- mediated triangles. 
For example, the tree level coupling
$a_L^e = t_3^e - Q_e \sin^2\theta_W$ of $Z$ to the left-handed electron is 
modified by $\sim f^2_{Ee}/16\pi^2 = 6.3 \times 10^{-5}$ for $f_{Ee} 
\sim 10^{-1}$. It is perfectly compatable with the precision 
of leptonic branching ratio of $Z$ at LEP which is 
presently at the per mille level. 
Flavor-violating $Z \rightarrow e\mu$ will also be induced for
simultaneous presence of $e$- and $\mu$-related new Yukawa couplings 
generating an effect of order 
$\sim \left(f_{Ee}f_{E\mu}/16\pi^2\right)^2$ and the condition of 
satisfying $\mu \rightarrow e \gamma$ automatically takes care of 
its consistency with experiment. 
The new Yukawa couplings also lead corrections to $g-2$ of electron
 of order $\simeq {{f^2_{e}m^2_e}\over{16\pi^2 M^2_{\eta}}}$
which is at the level of $10^{-15}$ for our choice of parameters
safisfying present measurements. 

\noindent(ii) The standard neutrinos are massive in this model. 
However, their masses
are arbitrary since they depend on the values of the corresponding Dirac
masses from the see-saw formula and hence can be tuned to the desired
values.

\noindent(iii) A recent publication by the L3 Collaboration of LEP
\cite{l3} gives experimental lower limits on the masses of 
the sequential leptons 
$E$ and $N$ from their non-observation. They exclude
the range $M_E < 61$ GeV and $M_N < 48.6$ GeV on the basis of nearly 
6 pb$^{-1}$ data collected at $\sqrt{s}$ = 130 -- 136 GeV run at LEP 
last year. Since we assume these
masses in the 100 GeV range, our model is consistent with these
bounds. The possibility of observing the sequential leptons in the 
oncoming phases of LEP2 run have been investigated \cite{bc} with 
the conclusion that their mass reach could go very close to their
kinematic limits under favorable conditions. 

\noindent(iv) It may be noted that the masses of the fourth generation
leptons are bounded by the electroweak symmetry breaking 
scale. As far as the neutrino states of the fourth generation are 
concerned, the masses of the two Majorana eigenstates are $M_{N_1}
\simeq v^2_{wk}/M$ and $M_{N_2} \simeq M$, induced by the see-saw
mechanism. The requirement that the lighter one is heavier than 
$M_Z/2$ (from the $Z$-invisible width constraint at LEP) implies an
upper bound  $M_{N_2} < 2v^2_{wk}/M_Z \simeq 1.3$ TeV on the heavier 
eigenstate \cite{zhang}. So future colliders, e.g. NLC have chances to
see them under favorable conditions. 

In conclusion, we have presented a non-supersymmetric interpretation of
the CDF $\eegg$ event by invoking new physics at the
electroweak scale in the context of an extended particle content 
for the SM that has a fourth sequential fermion generation
and massive Majorana right-handed neutrinos and a singly charged 
scalar. The kinematics of our model can be set exactly analogous to 
the SUSY scenario while fitting the CDF event -- the singly charged 
scalar playing the role of selectron and the right-handed neutrino 
acting as a counter-part of the next to lightest supersymmetric particle. 
We admit that our scenario is quite 
{\em ad hoc} and tailored to fit the CDF $\eegg$ event. However,
it has some features quite distinct from SUSY and, if this type of `zoo 
event' shows up in large number, it may be possible to distinguish 
between the two scenarios. 

\vspace{4mm}
{\bf Acknowledgement}
\vspace{4mm}

The work of R. N. M. is supported by a grant from the National Science 
Foundation and a Distinguished Faculty Research Award by the University
of Maryland. G. B. would like to acknowledge the hospitality of the
University of Maryland where most of the work was done and thank Riccardo 
Barbieri for a careful reading of the manuscript and suggesting 
improvements.


\end{document}